\documentclass[prl,aps,twocolumn,showpacs,floatfix]{revtex4}

\usepackage{epsfig}
\usepackage{amsmath}
\usepackage{amssymb}
\usepackage{amsfonts}
\usepackage{graphicx}
\usepackage{color}
\usepackage{bm}

\newcommand{\bb}{\begin{equation}}
\newcommand{\ee}{\end{equation}}

\begin{document}

\title{Density of states measurements for heavy subband of holes in HgTe quantum wells}

\author{A.~Yu.~Kuntsevich$^{a,b}$, G.~M.~Minkov$^c$, A.~A.~Sherstobitov$^c$, Y.~V.~Tupikov$^d$, N.N. Mikhailov$^e$, S.A. Dvoretsky$^e$}
\affiliation{$^a$Lebedev Physical Institute of the RAS, 119991 Moscow, Russia}
\affiliation{$^b$National Research University Higher School of Economics, Moscow 101000, Russia}
\affiliation{$^c$Ural Federal University,Ekaterinburg, Russia}
\affiliation{$^d$Department of Physics, Pennsylvania State University, University Park, PA 16802, USA}
\affiliation{$^e$Institute of Semiconductor Physics, Novosibirsk 630090, Russia}

\begin{abstract}
{Valence band in narrow HgTe quantum wells contains well-conductive Dirac-like light holes at the $\Gamma$ point and poorly conductive heavy hole subband located in the local valleys.}
Here we propose and employ two methods to measure the density of states for these heavy holes. The first method uses a gate-recharging technique to measure thermodynamical entropy per particle. As the Fermi level is tuned with gate voltage from light to heavy subband, the entropy increases dramatically, and the value of this increase gives an estimate for the density of states. The second method determines the density of states for heavy holes indirectly from the gate voltage dependence of the period of the Shubnikov-de Haas oscillations for light holes. The results obtained by both methods are in the reasonable agreement with each other. Our approaches can be applied to measure large effective carrier masses in other two-dimensional gated systems.
\end{abstract}

\maketitle

\section{Introduction.}


For two-dimensional (2D) system{s} the density of states $D$ is proportional to the effective mass of carriers $m_{c}$. 
Effective mass of carriers is usually measured either from {temperature} damping of quantum magnetooscillations or from cyclotron resonance.
{However,} both the mobility and the cyclotron splitting 
are {\it inversely} proportional to the effective mass. Graphene {with negligible electron mass} sustains quantum Hall effect up to room temperature \cite{graphene}, whereas semiconducting AIIIBV quantum wells (with $m_{c}\sim 0.1m_0$, where $m_0$ is free electron mass) {demonstrate magneto-oscillations} up to $\sim 10$ K\cite{ando}, and heavy fermion systems {require} well below 1~K\cite{heavyfermion}. {Therefore, for} heavy carriers the cyclotron splitting is small, their magneto-oscillations are damped, {and the density of states is not easily accessible.}  {Alternative approach of tunnel experiments\cite{Jang2017} requires sophisticated sample design and unprecedented quality of the tunnel barrier material.}
{

On the other side, {entropy and specific heat are }proportional to the thermodynamical density of states{, and hence the} effective mass. {These values should be} more accessible for heavy carriers. 
{Still it is challenging to measure } the entropy {and specific heat} for 2D system{, since they are} much less than for the substrate.

\begin{figure*}
\vspace{0.1 in}
\includegraphics[width=500pt]{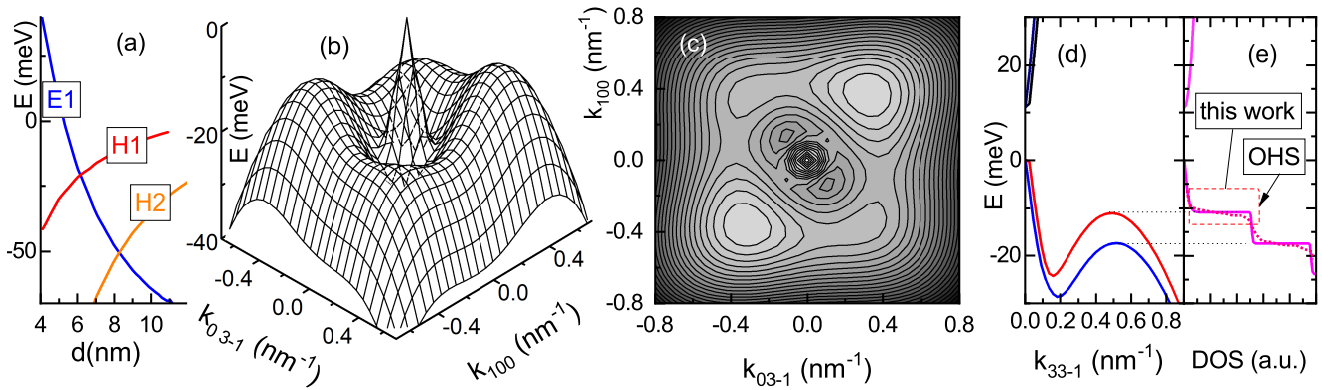}
\begin{minipage}{6.7in}
\caption{(Color online) {(a) QW thickness dependence of the relevant terms in Cd$_{0.7}$Hg$_{0.3}$Te/HgTe/Cd$_{0.7}$Hg$_{0.3}$Te quantum wells;} 3D view({b}) and isoenergy plot({c}) of the  valence band dispersion in 6 nm thick (0~1~3) Cd$_{0.7}$Hg$_{0.3}$Te/HgTe/Cd$_{0.7}$Hg$_{0.3}$Te quantum wells {(IIA is taken into account, see text)}(d) $E(k_{3 3 -1})$-dependencies demonstrating two spin-orbit-split branches of the dispersion law{; (e}) Corresponding density of states plot {(solid line), and the effect of disorder on OHS (dotted line)}.}
\label{fig1}
\end{minipage}
\vspace{0.4 in}
\end{figure*}

Strained HgTe quantum wells (QWs) open a vast playground for tuning the spectrum of 2D carriers as the width of the QW $d$ increases. {For the most common Cd$_{0.7}$Hg$_{0.3}$Te/HgTe/Cd$_{0.7}$Hg$_{0.3}$Te QW system t}he spectrum varies between ordinary gapped {one} ($d<6.3 nm$) \cite{ortnerVB, minkovVB}, graphene-like \cite{buttnerDirac} ($d\sim6.3$ nm), two-dimensional topological insulator\cite{konig} (6.3 nm$<d\lesssim 12$ nm), semimetal\cite{minkovsemimetal,kvonsemimetal} ($12<d\lesssim 30$nm), and even three-dimensional topological insulator in thicker films of strained HgTe ($d\sim 50 - 60$ nm)\cite{kozlov}. Remarkable property of the valence band spectrum for rather thin QWs is that besides Dirac-like carriers in the center of the Brilluin zone, formed by E1 and H1 subbands, it contains separated {local} valleys with large density of states that are formed by H2 subband (see Fig.\ref{fig1}a). As $d$ increases, H1 and H2 float up in energy, whereas the E1 subband moves down, and the Dirac feature at the $\Gamma$-point  eventually disappears for the semimetallic QW thickness. In the inverted band structure HgTe QWs for $d>8$~nm, the mass of heavy holes {(doubly degenerate by valley)} is measurable and reasonably small ($\sim 0.2 - 0.3 m_0$) \cite{minkovVB2017}. {In the narrower QWs with both inverted and normal spectrum,} quantum oscillations from the local valleys were not observed so far. Correspondingly there is no experimental information about {their} density of states. 

To fill {this} gap in the knowledge, in our study we measure the density of states {of the valence band} in the narrow HgTe QWs($<6.3$~nm) using two methods: (i) thermodynamical entropy-per-electron recharging technique and (ii) indirect measurements from Shubnikov-de-Haas density of light holes. The first method, developed by us in Ref.\cite{kuntsevich}, is sensitive to the variation of the density of states (averaged over $k_BT$ interval about Fermi energy) with total hole density. As a new subband starts filling, the entropy-per-electron value experiences a spike\cite{varlamov} with the amplitude directly proportional to density of states. The second method exploits the fact that total hole density of states is a sum of heavy and light hole densities. For the latter, both effective mass and density can be found from Shubnikov-de Haas oscillations (SdHO). The results obtained by two methods are consistent with each other within the measurement errors. Both methods can be further applied to 2D systems, consisting {of both} heavy and light carriers.

\section{Spectrum in narrow ${\rm {\bf {\large HgTe}}}$ QWs: present state}

In this section we review the present theoretical and experimental understanding of the valence band spectrum in the narrow Hg QWs. {The spectrum in QWs deviates from that in zero-gap bulk HgTe semiconductor due to size quantization effects and strain, which is caused by HgTe/Cd$_x$Hg$_{1-x}$Te lattice mismatch. The strain depends on $x$ (that is usually close to 0.7), and weakly depends on crystallographical orientation of the growth direction, usually (0~0~1) or (0~1~3). 

The bottom of the conduction band is located at the $\Gamma$ point.}
{For a wide range of $d$, the dispersion $E(k)$ in the valence band  has a
central extremum at the $\Gamma$-point and four local maxima in the valleys. 
For $4<d<7$ nm, the top of the valence band is located at the $\Gamma$ point and has almost isotropic dispersion.
For $d>8$ nm, the local valleys float up in energy above the {central} extremum (see H2 term in Fig.\ref{fig1}a).

For hypothetical HgTe QWs with symmetric barriers and symmetric potential profile in $z$-direction, the dispersion law is two-fold degenerate (by spin). In the most widespread structures with
orientations (1~0~0) and (0~1~3), local maxima in $k$-space form a
square and a rectangle, respectively. Therefore, in this approximation, the local valleys should have 8-fold degeneracy (4 valleys $\times$ 2 spins).

The absence of a center of inversion in  parent materials having zinc-blende structure leads to asymmetry of the interface (i.e. interface  inversion asymmetry or IIA), which causes
spin-orbit splitting of the spectrum and lifts the two-fold spin degeneracy. 
Correspondingly, the states near the central extremum split into two branches, and the local valleys near their tops become 4-fold degenerate.
In the framework of $(k\cdot p)$ theory, the contribution of IIA is determined by the
phenomenological parameter $g_4$ \cite{tarasenko}. Note, the barriers and potential profile in $z$-direction are still assumed to be symmetric here.


Additional asymmetries such as
difference in the widths of the
heteroboundaries, non-zero average electric field in the QW, and different values of the parameter $g_4$ at the top and bottom borders of the QW, lead to the removal of this 4-fold valley degeneracy. Local valleys become 2-fold degenerate near their tops both for (1~0~0) and (0~1~3) substrate orientation (see Fig.\ref{fig1}b,c).

Experimentally, for reasonably low carrier density $p<(3-6)\cdot 10^{11}$ cm$^{-2}$, the twofold valley degeneracy in HgTe QWs with $d>8$~nm was indeed demonstrated\cite{minkovVB2017}.
 Effective mass in the valleys weakly depends on $d$ and grows with $p$ (from $\sim 0.2 m_0$ to $0.3 m_0$)\cite{minkovVB2017}.
It was shown in Ref.~\cite{minkovVB}, that the value of the parameter 
$g_4 = 0.6-0.8$~\AA eV reasonably explains the experimentally observed masses and spin-orbit splitting in the central extremum.
A bit larger values of $g_4= 0.8 - 1$ \AA eV describe spectrum of local valleys \cite{minkovVB2017}.
}

In Figs. \ref{fig1}b,c we present the results of {the} spectrum calculation within four band Kane model with IIA for 6 nm thick (013) Cd$_{0.7}$Hg$_{0.3}$Te/HgTe/Cd$_{0.7}$Hg$_{0.3}$Te quantum wells, using the parameters from  Ref.\cite{novik}, and IIA parameters {for top and bottom interfaces of the QW} $g_{4top}=0.8$ \AA eV, and $g_{4bottom}$=1 \AA eV{, respectively. The top valleys are two-fold-degenerate within this model.}

 For the narrow HgTe QWs,  density of states has a jump at the onset of heavy subband (OHS) point. {This cusp should be } slightly smeared by sample disorder, as shown in Fig.~ \ref{fig1}e. The amplitude of the cusp is proportional to the effective mass and the total degeneracy of the {top} valleys. 

Real spectrum of the local valleys for $4<d<7$ nm QWs, however, remains unknown {because n}o information (effective mass, degeneracy and thus density of states) about these heavy subbands was available so far. {These valleys did not reveal} SdHO for hole density up to $p=6\cdot 10^{11}$ cm$^{-2}$ in magnetic fields up to 7 T and temperatures down to 1.4K.

\section{Samples}
We studied several samples made of strained QWs grown on (013) GaAs substrate (wafers 110622 with $d=5.6$ nm , and 110623 with $d=6$ nm, that corresponds to normal spectrum). All QWs were p-type  at zero gate voltage. First, mesas were defined lithographically,
then the samples were CVD covered with 300-400 nm
parylene insulator, and finally  100~nm aluminium {top-gate was thermally evaporated}. The structure of the sample is shown in Fig.\ref{fig1a}a. We defined large 4$\times$5 mm$^2$ mesas for thermodynamic measurements, and standard Hall bar mesas with inter-contact distances 0.5 mm for transport measurements.

\begin{figure}
\vspace{0.1 in}
\includegraphics[scale=0.3]{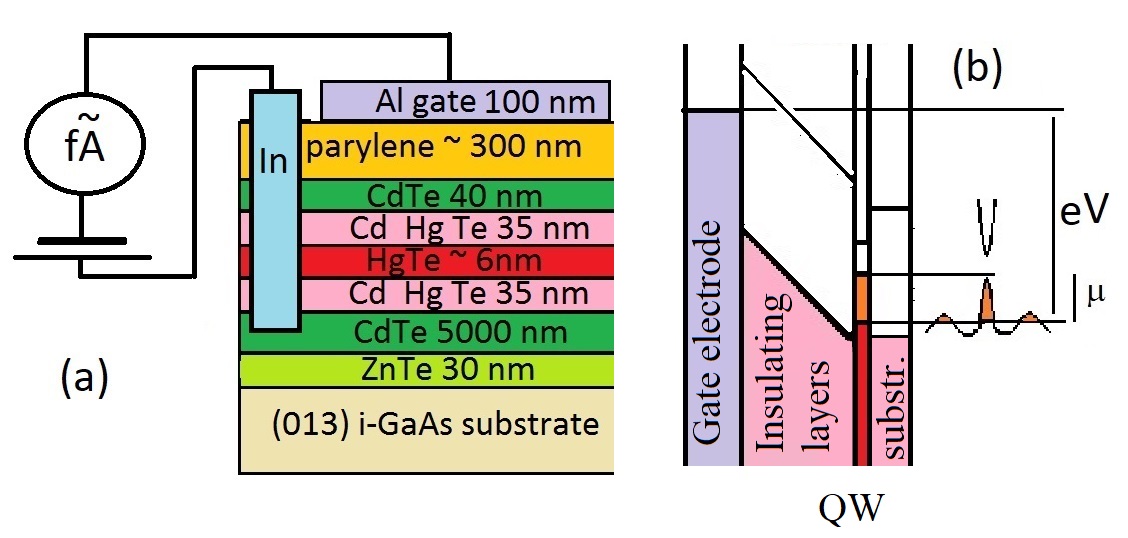}
\begin{minipage}{3.2in}
\caption{(Color online) (a){Gate-insulator-QW structure and schematic for entropy measurements. Ammeter measures recharging femtoampere-range current in response to temperature modulation.}; (b) band diagram of the Gate-insulator-QW structure{, showing holes (orange) in the QW. Dispersion law for the holes is indicated to the right from the band diagram.}}
\label{fig1a}
\end{minipage}
\vspace{0.4 in}
\end{figure}
{\section{Entropy detection of heavy holes: qualitative discussion  and measurement scheme}
}

Let us first consider entropy {of the degenerate Fermi gas} for $T\rightarrow 0$. 
Independently of {the spectrum and} dimensionality, {the entropy} is proportional to number of excitations near the Fermi energy {($\epsilon_{F}\equiv\mu$ at $T\rightarrow 0$)} , i.e. to {the temperature $T$} and the density of states $D$ at the Fermi level:
\begin{equation}
S=\pi^2 TD/3
\label{entrop}
\end{equation}
In a perfect no-disorder sample at the OHS point, entropy would change with density stepwise, and hence one would expect $\partial S/\partial p\propto \delta (\epsilon_{F} - \epsilon_{\rm OHS})$. Disorder{, temperature,} and sample inhomogeneities naturally broaden {the }$\partial S/\partial p(\epsilon_{F})$ dependence. 

Temperature dependence of $\partial S/\partial p$ is determined by the interplay of two factors: (i) the number of excitations, {that} increases with $T$ (see Eq.\ref{entrop}), (ii) non-degeneracy at high temperatures, i.e. emergence of the Fermi energy scale, that restricts the entropy growth.

{A method for entropy measurements has been recently developed for the field effect structures \cite{kuntsevich,tupikov}.}  The method exploits the fact that the voltage between QW and the gate electrode consists of electrostatic ($E\cdot d$) and thermodynamic($\Delta \mu/e$) parts{
, as shown in the band diagram in Fig.\ref{fig1a}b. 

To modulate the temperature, we heat resistively a container with a sample and a thermometer, that is connected to the cryostat through the heat exchange gas. Heating power, hence the temperature, is modulated at double frequency, which allows us to detune from the parasitic first harmonic signals using the lock-in technique. In response to modulation of temperature $T(t)=T_0+\Delta T \cos(\omega t)$, the chemical potential oscillates $\Delta \mu (t)=\Delta\mu _0+\partial\mu /\partial T \cdot \Delta T \cos(\omega t)$. Since the gate voltage is maintained constant, $E(t) \cdot d +\Delta\mu (t)/e=const$.  This leads to the out-of-phase femtoampere-range current $j$ through the capacitor:}
\begin{equation}
j(t)=\frac{\omega C}{e}\frac{\partial\mu}{\partial T}_{p}\Delta T \sin (\omega t)
\label{modcur}
\end{equation}

 By measuring {this} recharging current {with ammeter, shown in Fig.\ref{fig1a}a,} we straightforwardly determine the temperature derivative of the chemical potential. This derivative through the Maxwell relation
\begin{equation}
\partial S/\partial p_{~T}=-\partial \mu/\partial T_{p},
\label{maxwell}
 \end{equation}
 allows us to measure derivative of entropy with respect to {the hole} density. In Eq.\ref{maxwell} we use chemical potential for holes, that has an opposite sign to the one of electrons. By integrating this derivative with hole density we evaluate the variation of the entropy.  {This method was first applied to} archetypical Si-MOSFET and GaAs/AlGaAs heterojunction two-dimensional systems\cite{kuntsevich, tupikov}, with a simple quadratic dispersion. The non-zero signal in these measurements might come from non-degeneracy, or from electron-electron interactions, or from Landau quantization in perpendicular magnetic field\cite{tupikov}. On the contrary, for HgTe QWs the electron-electron interactions are weak and the entropy-per-electron signal is expected to result from the emergence of the heavy subband.

We also note that at the charge neutrality point the density of states reaches minimum and therefore $\partial S/\partial p$ should change sign.
 
{\section{Capacitance and entropy detection of heavy holes: experimental}
}
 {It was demonstrated previously (see discussion of Fig. 2a and Fig.1(f,g) in Ref.\cite{kuntsevich}) that the entropy method in archetypal 2D system in Si-MOSFET gives a reasonable value of effective mass ($\sim0.2m_0$) and cyclotron splitting. Thus, we expect the method to produce valid results for HgTe QWs.}  
 {In order to find $\partial \mu/\partial T$ from Eq.~\ref{modcur}, and to identify CNP, one has to know the capacitance $C$.  We determined the capacitance by modulating the voltage and measuring the recharging current.  We used the same room-$T$ operated current-voltage converter as for current-voltage converter as for $\partial\mu/\partial T$.} 

 The capacitance $C$ of the gated QWs is known to consist of two contributions, connected in series:
\begin{equation}
\frac{1}{C}=\frac{1}{C_{geom}}+\frac{1}{C_q}
\end{equation}
geometrical one $C_{geom}=A\epsilon_0\epsilon_{eff}/d_{eff}$ and quantum one $C_q=Ae^2\partial p/\partial \mu$. Here $A$- is the sample area, $d_{eff}$ is effective gate-to-2D gas distance, $\epsilon_{eff}$ is effective dielectric constant of the insulating layer. If one neglects fine correlation effects \cite{quantcap} and effects of finite temperature, one gets $C_q=Ae^2 D$. The qualitative difference between capacitance and $\partial S/\partial p$ is that the latter is sensitive to density of states and may be used to detect heavy carriers, whereas the capacitance is sensitive to inverse density of states and used to detect gaps in spectrum\cite{khrapai} or light carriers\cite{kozlov2}. As a rule, quantum capacitance much exceeds the geometrical one, and $C_q$ becomes distinguishable only close to charge neutrality point, where $D\rightarrow 0$. For our large area samples, OHS feature in capacitance was smeared, and could not be identified on the top of the geometrical capacitance. Moreover , the geometrical capacitance always has some weak {$V_g$-}dependence. {In single QW structures it is impossible to fully disentangle geometrical and density-of-states contributions}. Possible solution would be to use double-QW structures with separate contacts to each layer (similarly to Ref.~\cite{eisenstein}) to explore the compressibility. The latter method naturally compensates the geometrical contribution, however sample production becomes very tricky, as HgTe QWs do not survive treatment above 100 $^\circ$C. Therefore, $\partial S/\partial p$ provides an alternative way to access the density of states, which is superior to the capacitance measurements for our QW samples.

\begin{figure}
\vspace{0.1 in}
\centerline{\includegraphics[width=250pt]{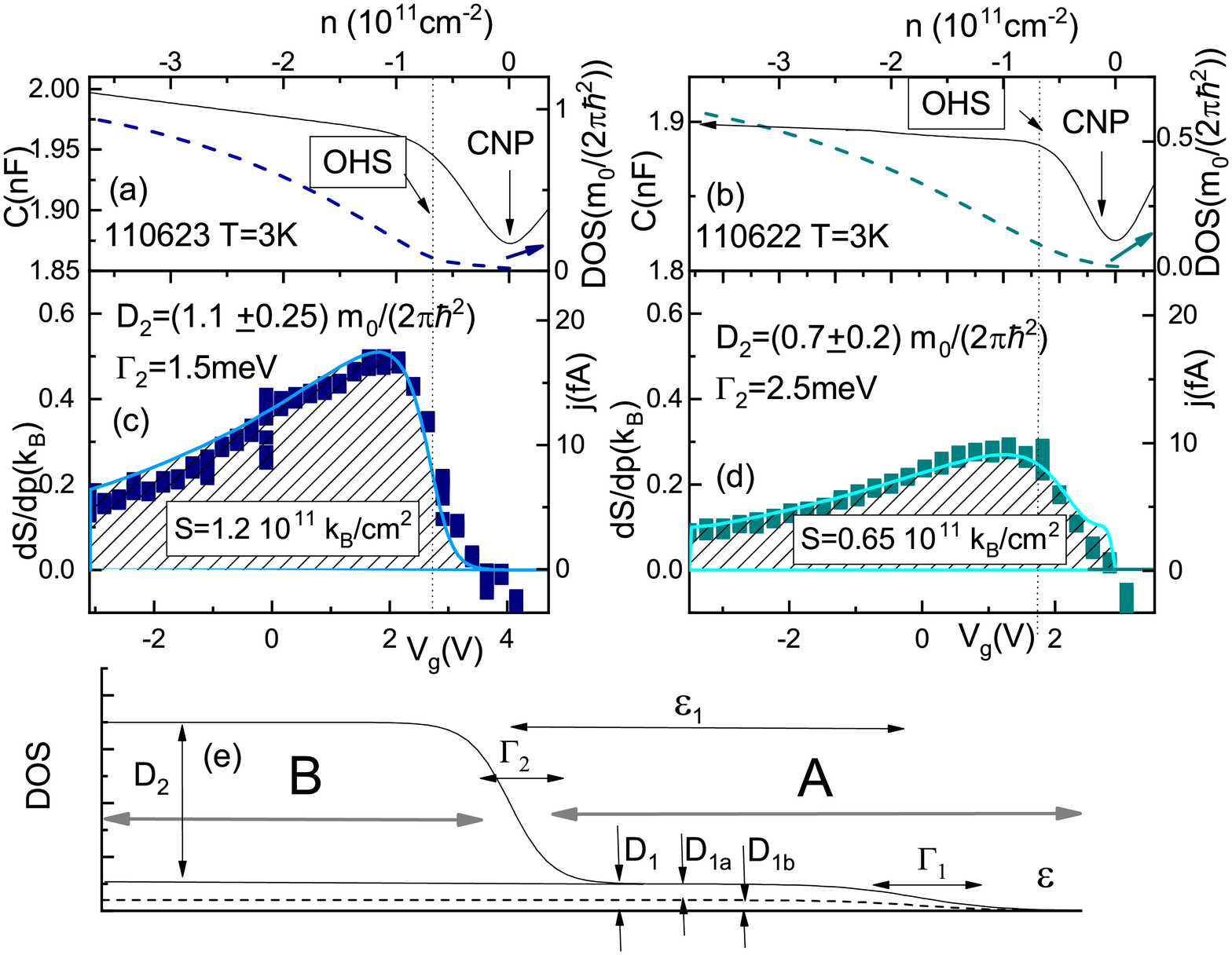}}
\begin{minipage}{3.2in}
\caption{(a,b) Capacitance of the samples 110623 and 110622 versus carrier density measured at the same measurement frequency as $\partial S/\partial p$ {(solid lines, left axes), density of states at the Fermi level, recalculated using the model for $\partial S/\partial p$ fit (dashed lines, right axes)}; (c,d) $\partial S/\partial p$ value for the same samples versus density at 3K. Lines - theoretical fits using smeared steplike density of states model. Hatched areas show the integrated entropy value for $p=3.7\cdot10^{11}$~cm$^{-2}$. {Right axes in panels (c,d) show the root-mean-square value of the recharging current.} (e) Density of states model and parameters used for fitting of the data in panels b and d. Gray arrows denote two regions (A, where only light carriers are present, and B where also heavy subbands emerge)}
\label{Fig2}
\end{minipage}
\vspace{0.4 in}
\end{figure}

In our experiments, the temperature was varied from 3 to 20 K. For the reasons, explained in Appendix I, we believe that the reliability of the results decreases with temperature, and therefore consider only the lowest temperature 3K.
{Root mean square temperature modulation $\Delta T$ was about 0.05~K, and frequency $\omega/(2\pi)$ was about 0.7 Hz.}
Examples of experimental $\partial S/\partial p (p)$ data for the samples 110623 and 110622 are shown in Fig. \ref{Fig2}(panels c and d, respectively). Capacitance measurements performed in the same samples at the same frequencies clearly indicate minimum at CNP (Fig. \ref{Fig2}a,b). One can see that $\partial S/\partial p$  signal increases at about $p\sim 10^{11}$ cm$^{-2}$, in-line with saturation of the Hall resistance at the OHS point(see next section). Moreover, at the charge neutrality point $\partial S/\partial p$ changes sign, as expected.
Together these observations support the reliability of $\partial S / \partial p$ data recorded at 3K.

The exact spectrum in HgTe quantum wells remains {not fully understood}\cite{minkovVB,minkovsemimetal}. 
In addition, its temperature dependence\cite{krishtopenko}, and the role of interaction effects are not yet clarified \cite{quantcap}. 
Thus, we decided to model the density of states with a simple smooth step-like function. {According to our measurements in thicker QWs \cite{minkovVB2017}, the dispersion in local valleys is quadratic. This suggest constant density of states in heavy hole subbands.} Figure \ref{Fig2}e illustrates model density of states, that is determined by  five parameters: densities of states below ($D_1$) and above($D_1+D_2$) OHS, broadening of the band tails($\Gamma_1$,$\Gamma_2$), position of the OHS in energy $\epsilon_1$. Mathematically, we approximate density of states near the OHS in the form $$D(\epsilon)=D_1+0.5\cdot D_2\cdot [1+tanh({\Gamma_2}^{-1}\cdot(\epsilon-\epsilon_1))]$$

{Under the condition $D_1\ll D_2$ (confirmed by numerical checks) the $\partial S/\partial p (p)$ dependence, produced by this ansatz, is sensitive mainly to $D_2$ and  $\Gamma_2$.  The product $D_1\cdot\epsilon_1$ sets the hole density at OHS point, and can be reasonably well estimated from the Hall effect data(see next section).

Therefore, we adjust two parameters (density of states for heavy carries $D_2$\cite{aboutchempot}, and the width of the bandtail $\Gamma_2$) to obtain the best fit of $\partial S/\partial p(V_g)$ data. More specifically, we do the following. W}e use the definition of the carrier density:

\begin{equation}
p(\mu)=\int_0^\infty\frac{D(\epsilon)}{1+\exp[(\epsilon - \mu)/T]} d\epsilon
\label{totalcharge}
\end{equation}

 Next we solve this integral equation with respect to $\mu$. Taking derivative of $\mu(T)$ dependence we determine $\partial S/\partial p=-\partial\mu/\partial T$.
By varying $\Gamma_2$ and $D_2$, we achieve the best agreement with the experiment (lines in Fig.\ref{Fig2}b,d). 
Using $D_2$, one can estimate the effective mass, provided that degeneracy of the heavy subband is known. In Fig. \ref{Fig2}c,d we give the values of effective masses, as if these holes were not degenerate.  Assuming the twofold heavy hole degeneracy(see discussion), the effective mass for 110623 is about $(0.55\pm 0.12) m_0$ and for 110622 is about $(0.35 \pm 0.1)m_0$.
We evaluate the errors of $\partial\mu/\partial T$  measurements (and hence the proportional value $D_2$) as $\sim 25$\%. This value reflects mainly the uncertainty in the temperature modulation depth throughout the measurement cell (possible differences of temperature at the thermometer and at the sample).

The values of $\Gamma_2$ (1.5~meV for 110623 and 2.5 meV for 110622) are rather huge: the typical amount of the carriers within the bandtail $\Gamma_2D_2/2$ is about the OHS density. Interestingly, similar values of the band-tails were observed in thicker HgTe QWs in Ref.~\cite{kozlov2} from capacitance measurements. In our large area samples, this value of $\Gamma_2$ might partially reflect the nonuniformity of the gate thickness over the sample area. 
\begin{figure*}
\vspace{0.1 in}
\centerline{\psfig{figure=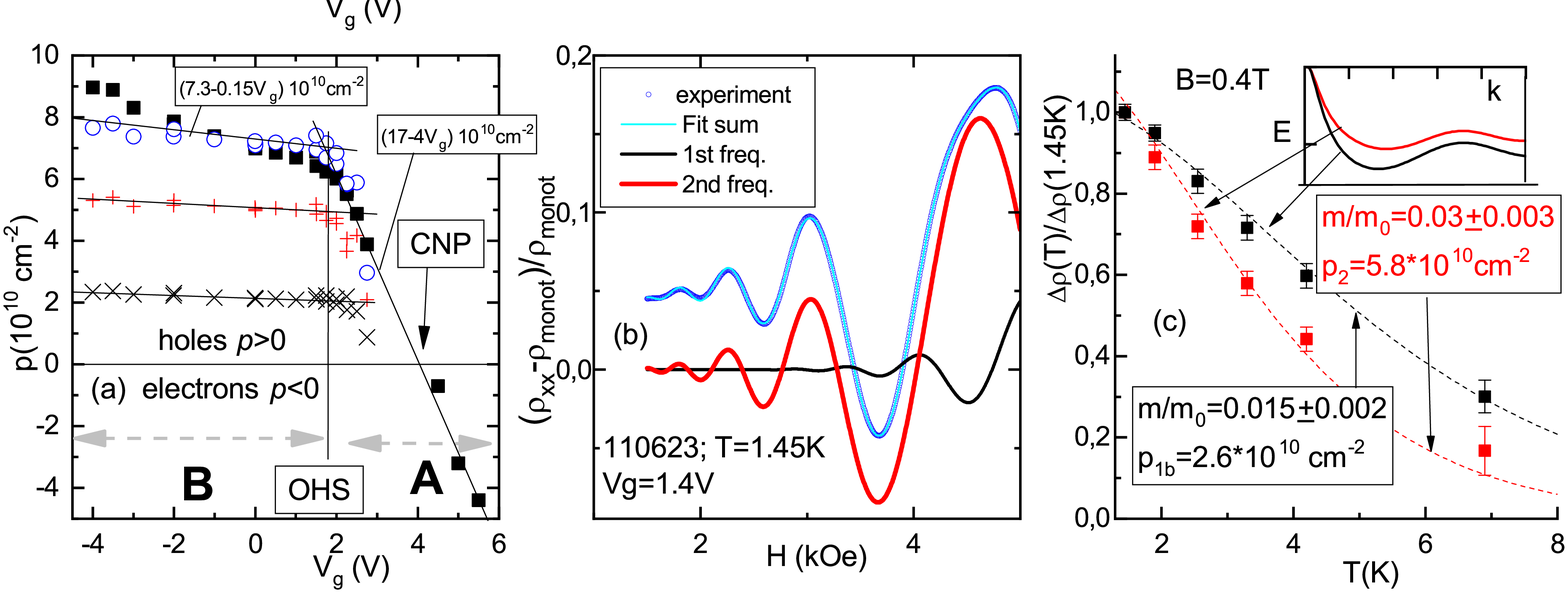,width=500pt}}
\begin{minipage}{6.7in}
\caption{(a) Carrier density of light subbands (Crosses), determined from ShDH oscillations frequencies and their sum (empty circles) versus gate voltage. Solid boxes - carrier density determined from Hall constant at 0.2~T. Lines show the corresponding linear $p(V_g)$ fits.(b) Example of SdHO pattern for sample 110623 and their expansion into two frequencies, corresponding to two spin-orbit split dispersion branches.  (c)Dingle plots for both branches and effective mass determination. }
\label{fig3}
\end{minipage}
\vspace{0.4 in}
\end{figure*}

{It is instructive to plot the gate voltage dependence of the density of states at the Fermi level in the model (dashed lines in Fig.~\ref{Fig2}a,b). The increase of $D$ with $p$ by 1.5 orders of magnitude 
looks more gradual compared to the abrupt jump in
$D(\epsilon)$ dependence. This is because the larger the density of states, the slower Fermi energy changes with gate ($\delta V_g\propto \delta p \approx\delta \mu /D)$.}

\section{Transport detection of heavy holes}

We started by measuring a Hall density ($p_{Hall}=e^{-1}{R_H}^{-1}$) dependence on the gate voltage. The emergence of the heavy subband in the Hall data is revealed by a drastic change in the $dn/dV_g$ slope at $V_g\sim$2~V (see black squares in Fig. \ref{fig3}a). It is seen, that for $3{\rm V}<V_g<6{\rm V}$,  Hall density changes linearly with $V_g$, $dp/dV_g=-0.4\cdot 10^{11}$ cm$^{-2}$V$^{-1}$ (domain A). At the charge neutrality point (CNP) the sign of carriers changes. 

 For $V_g\lesssim 2$ V (domain B), Hall density becomes much less sensitive to the gate voltage. 
Such behaviour results from the pinning of Fermi level by some states with high density of  states{\cite{aboutchempot}. Small growth of $dp_{Hall}/dV_g$ for $V_g <-2$~V  is related to contribution of the local valleys to conductivity (see Appendix II).}
 
 Indeed, Hall density is determined by high mobility light holes, whereas heavy carriers are weakly conductive, and almost do not contribute to the Hall effect. At the same time, their density of states is much larger, and therefore mostly heavy states from local valleys are filled in with decreasing the gate voltage. In other words, heavy carriers lead to pinning of the Fermi level\cite{minkov2016}.	 Absence of SdH oscillations might originate either from elevated value of the effective mass or from stronger scattering. 

Multi-liquid model is a common approach to identify the densities and mobilities of different carriers from nonlinear magnetoresistance and Hall effect.
In our case however, precision of such measurements is not high as three groups of carriers are involved (two branches of light carriers and heavy subband). Low-temperature data are also superimposed with Landau quantization effects. In Appendix II we show that multi-liquid model at elevated temperatures allows to determine the density and mobility of low mobility holes and a sum of densities of both high mobility groups of carriers at the central extremum.  Extracted mobility for low mobility carriers is high, so their conductance exceeds $e^2/h$. That corresponds to band (delocalized) states. Although non-linear Hall effect and magnetoresistance are consistent with the low temperature transport and entropy data, they do not add any new information about masses or degeneracies.

A straightforward is to measure the density of light holes precisely through SdHO. While we observed no SdHO for heavy holes, the oscillations for light holes are very prominent and allow us to determine the density of light holes $p_l$ from SdHO frequency.   Knowing the $p_l$ value at $V_g$ close to OHS and density of states for light carriers, we can evaluate the density of states for heavy carriers.

The ratio of the slopes $k=dp_1/dV_g$ in the domains B and A (Fig.\ref{fig3}a) is given by the ratio of densities of states (see Fig.\ref{Fig2}e):
\begin{equation}
D_2=D_1\cdot (k_A/k_B-1)
\label{D1D2}
\end{equation}

Derivation of this formula is given in Appendix III.
Intuitively, due to the common Fermi level for light and heavy holes when both carriers are present, introduction of a new hole from light subband will add up $D_2/D_1$ holes from the heavy subband.

To determine the density of states in local valleys $D_2$, according to Eq.~\ref{D1D2}, one should not only know the $k_A/k_B$ ratio, but also density of states $D_1$ close to OHS point. $D_1$ is given by effective mass, that in turn can be found from the temperature dependence of the SdHO amplitude.

Determination of effective mass is hindered by the fact that central extremum of the dispersion law is split by spin-orbit interaction into two branches with different masses\cite{minkovVB}. 
To disentangle these masses, we fitted the magnetooscillations $\rho_{xx}(B)$ with a sum of two Lifshits-Kosevich formulas. The result of fitting for sample 110623 at $V_g=1$~V, and $T=1.4$~K is shown in Fig.~\ref{fig3}b. 
Despite many fitting parameters ($p_1$, $p_2$, parameters of damping $\delta_1$, and $\delta_2$, prefactors and phases), this procedure is superior in precision over the commonly used Fourier analysis. {In Appendix IV we consider this fitting procedure and demonstrate that it gives reproducible (with respect to the change in magnetic field range) values of frequency and amplitude for both branches.}

We used similar decomposition at various temperatures and gate voltages. 
In the domain A in Fig.\ref{fig3}a the heavy carriers are absent.  Hall density expectedly coincides with the sum of non-degenerate SdHO densities of both species of light holes.
In the domain B, Hall density systematically exceeds the total SdH density due to some contribution of the local valleys to the conductivity.

Fig.~\ref{fig3}c shows the temperature dependencies of the oscillation amplitudes for both branches, taken at magnetic field 0.4~T. Standard Lifshits-Kosevich dependence $A(B,T)\propto Y(B,T)/sinh(Y(B,T))$, where $Y=2\pi^2k_BTm/(e\hbar B)$, well describes the experimental data with $m_1=(0,015\pm 0.002) m_0$  and $m_2=(0,03\pm 0.003) m_0$ values for two branches, respectively. That gives the total density of states in the central extremum $D_1=D_{1a}+D_{1b}= (0.045 \pm 0.005)m_0/(2\pi \hbar^2)$.

Using Eq.\ref{D1D2}, we evaluate the density of states in the local valleys: $D_2=k_a/k_b\cdot D_1 =(27\pm 7) \times (0.045 \pm 0.005)\times m_0/(2\pi \hbar^2)=(1.2\pm 0.3) m_0/(2\pi \hbar^2)$. This value is in reasonable agreement with the entropy measurements $(1.1\pm 0.25) m_0/(2\pi \hbar^2)$.

Similar analysis of the SdHO data for 110622 sample gives $m_1=0.022\pm 0.003$,$m_2=0.034\pm0.005$, $k_a/k_b=25\pm 5$, and, correspondingly $D_2=(1.4\pm 0.5)\cdot m_0/(2\pi \hbar^2)$. Similarly to entropy measurements, $D_2$ for 110622 sample is also visibly smaller than for 110623. Numerical agreement with thermodynamical data $(0.7\pm 0.2) m_0/(2\pi \hbar^2)$ is not good, yet acceptable.

The consistency of density of states obtained by two different methods is the central result of our paper, demonstrating the experimental accessibility of the density of states in the heavy subband.

	The value of $D_2$ allows us to determine the effective mass in the local valleys, provided that the degeneracy of the spectrum is known. If we believe that the degeneracy is 2, similarly to QWs with $d>8$ nm \cite{minkovVB}, then the obtained density of states (average value of two methods) will give  $m_2\approx0.6 \cdot m_0$ for 110623 sample. This value is 2-3 times larger than in QWs with  8~nm $<d<$20~nm \cite{minkovVB}($m_2=0.23-0.3 m_0$) . It is a puzzle why the mass of holes from local valleys is so large.

Alternatively, the elevated density of states could be explained by a larger valley degeneracy of 4 (instead of 2). However, the possible mechanism behind the emergence of additional degeneracy is not clear, since IIA parameters for the HgTe QWs{, which determine the degeneracy within four band Kane model, were previously measured} experimentally\cite{minkovVB2017}.

\section{Discussion.}
Thus, we provided estimates of the density of states for heavy holes in narrow HgTe QWs (i) from entropy per electron measurements and (ii) from the magnetooscillations of the light holes. The first method gives the absolute value of density of states, according to Eqs. (\ref{maxwell}), (\ref{modcur}) and (\ref{entrop}), because the samples'  area, $\Delta T$, and $\omega$ are known. This method is sensitive mostly to derivative $d D/d\epsilon$. The second method, on the contrary, is sensitive to D: it measures the ratio of densities of states for heavy and light carriers. In order to find the latter, we exploit SdHO and Lifshits-Kosevich formula and also the fact, that light hole states are spin-orbit splitted. The consistency of both methods signifies that our model of density of states (Fig.\ref{Fig2}e) is adequate.

Note that the spectrum of the HgCdTe/HgTe/HgCdTe QWs is, in general, $T$-dependent, because HgTe and HgCdTe have different thermal expansions\cite{krishtopenko}. Since in the quantum well the total carrier density is fixed, the evolution of the spectrum with $T$ should cause the drift of the chemical potential. This parasitic effect is added to $\partial \mu/\partial T$ signal. Therefore, Maxwell relation Eq.~(\ref{maxwell}) apparently becomes violated. 

How is it possible for the thermodynamic equality to be invalid?
The answer is related to thermal expansion. Indeed, the derivative $\partial \mu/\partial T$ in Eq.(\ref{maxwell}) is taken under constant volume constrain. However, if there is a thermal expansion of the lattice, the volume of the 2D gas changes  with $T$.
In the low-$T$ limit, the thermal expansion tends to zero, while at high temperatures it could {significantly} affect the results. 

The OHS point, where the density of states changes dramatically, could also be interesting for observation of the other novel effects.
For example, we expect that across the OHS point the phase of the SdHO should change. The mechanism of this change is thermodynamical, as explained in Ref.~\cite{berry}. Indeed, if all holes are light, the total density is conserved $p=const$, and the positions of the Landau gaps (resistivity minima) are given by Landau level degeneracy ($p=NeB/h$). 
When the local valleys are filled , the chemical potential becomes pinned. Therefore $\mu=const$ condition is applicable to light holes instead of $p=const$.
In this regime, the phase of the oscillations becomes sensitive to Berry phase, energy dispersion etc. However, the observation of this phase change is tricky because of presence of two spin-split branches of light holes.

Some other properties, like energy and spin relaxation, dephasing, electron-electron correlation, magnetic susceptibility, and plasmonic effects should also be revealed at the OHS. This plethora of phenomena remains largely unexplored both theoretically (due to complicated Hamiltonian of the problem) and experimentally.  Therefore, HgTe QWs give a vast platform to explore new physics.

We also note, that emergence of the OHS subband in density of states is not limited to the HgTe quantum wells. Recently the system with similar density of states versus energy profile was explored in InSe/graphene layered heterostructure\cite{kudrinsky}, where the Fermi level was tuned from light graphene carriers to heavy conduction band of InSe. Very similar effects of Fermi-level pinning were observed in epitaxial graphene on graphite\cite{graphenegraphite}, where light carriers of the top graphene layer were superimposed with low-mobility graphite large-density of states reservoir. We also note that all 3D topological insulators have extraordinary huge density of states for bulk 3D states, compared to 2D surface states. If for some reason Fermi level of the topological surface states touches the bottom of the conduction band, this is fully analogous to OHS phenomenon discussed in this paper\cite{berry}.  

Note that the subband should not necessarily be too heavy for observation of singular entropy-per-electron behaviour\cite{varlamov, germanene, diracentropy}. This implies applicability of the described techniques to the other material systems.

\section{Conclusions.}
We have measured the density of states in local hole valleys in narrow HgTe QWs ($d\lesssim 6.3$ nm, with non-inverted spectrum) using two different methods: absolute entropy per electron measurements and relative SdHO pattern analysis. The results of both methods are consistent with each other and give the total density of states, that corresponds to $m\approx 0.5 - 0.9 m_0$, assuming that local valleys are doubly degenerate. This observation is puzzling, as this value is 2$-$3 times larger than in QWs with $d>8$ nm, whereas band structure calculations predict only 10-20\% growth of effective mass when QW width decreases from 8 to 6 nm. 

\section{Acknowledgements.}

The authors are thankful to S.S. Krishtopenko, and I.S. Burmistrov for discussions. The measurements facilities of the LPI were used for entropy detection of the heavy carriers. These measurements were supported by Russian Science Foundation (Grant No. 17-12-01544). The transport measurements have been supported in part by the Russian Foundation for Basic Research (Grant No. 18-02-00050), by Act 211 Government of the Russian Federation, agreement No. 02.A03.21.0006, by the Ministry of Education and Science of the Russian Federation under Project No. 3.9534.2017/8.9, and by the FASO of Russia (theme "Electron" No. 01201463326). {AYK was supported by Basic research program of HSE.}

\section{Appendix I}
There are several mechanisms, that decrease the reliability of the entropy data at high $T$. Here we discuss how these parasitic mechanisms affect the $\partial S/\partial p$ signal and show that the data measured at elevated temperature contains unphysical regions, unlike the lowest-$T$ data. 

Increase of the container heat capacity with temperature, reduces the available amplitude and frequency of the temperature modulation, and hence the sensitivity of the method. The 
leakage current through the coaxial wires and parylene gate also tends to increase with $T$, thus restricting the range of the sensitivity to $T<20$ K.

The measured recharging current is proportional to ${\partial\mu}/{\partial T}$, sample area and modulation frequency. 
The latter is limited to below $\sim 1$ Hz due to finite thermalization time of a container (see details in Supplementary Information to Ref.\cite{kuntsevich}).
For sample capacitance $\sim 1$~nF and $\partial S/\partial p \sim k_B$, the {signal current is in} the femtoampere range (right axes in Fig.\ref{Fig2}c,d).
Therefore, to increase the current one has to increase the
capacitance, i.e., to take samples with large area (several square millimetres). Gradients of gate thickness, QW thickness, and doping in large samples are bigger . Therefore, thermodynamical features
become 
less singular.

We believe this large area effect is partially responsible for the difference between thermodynamical(obtained in large samples) and transport(obtained locally in small samples) values of $D_2$.

 Sample-size independent limitations 
 should also be discussed here. First, each capacitive structure has a leakage current, which is strongly nonlinear (usually threshold-type) function of gate voltage. 
The phase of this parasitic current, contrary to the phase of the signal, should coincide with the temperature modulation, thus allowing one to detune from it. Our practical observation is, however, that this leakage current is extremely noisy. As it achieves picoampere range, the measurements of $\partial \mu/\partial T$ become impossible. At the same time, this pA-range leak does not impede the nA-range capacitance measurements in the same sample in wide range of $p$.

\begin{figure}
\centerline{\psfig{figure=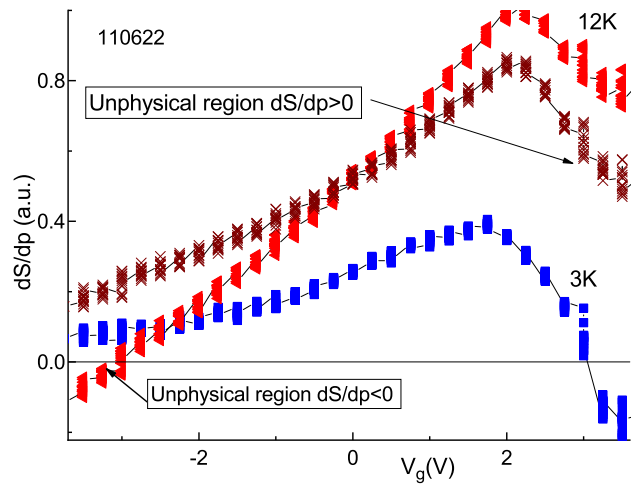,width=250pt}}
\caption{ Gate voltage dependence of the $\partial S/\partial p$ signal for sample 110622, at 3~K (blue squares) and 12 K, with (crosses) and without(triangles) subtraction of linear-in-$V_g$ correction(see text).}
\label{dsdnlimit}

\end{figure}

Second type of the sample imperfection is the temperature dependence of the geometric capacitance. This dependence originates either from heat expansion of insulating layer or from $T$-dependence of the dielectric constant of the insulating material. This parasitic contribution current $i_{par}$ is in-phase with the useful signal, and is equivalent to diamagnetic shift to $\partial\mu/\partial B$\cite{reznikov,teneh}:

\begin{equation}
i_{par}= \omega\Delta T V_g dC/dT \sin (\omega {t})
\label{cvst}
\end{equation}

It means that even if the variation of the capacitance is only a few ppm per K, for $V_g$ about few volts one gets parasitic current that corresponds to $\partial\mu/\partial T\sim 0.1 k_B$, i.e. of the same order as useful signal (the maximal possible value of $\partial S/\partial n$ 
is $\sim k_B$\cite{galperin}). At $\sim 3$ K we  do not observe temperature dependence of the capacitance, whereas as temperature increases, $dC/dT$ also increases, making measurements unreliable already for $T\sim 10$~K. 

Figure \ref{dsdnlimit} compares $\partial S/\partial p(V_g)$ data collected at 3~K(squares), that we believe are free from side effects, with 12~K data (triangles). The latter demonstrate unphysical values $\partial S/\partial p<0$ for $V_g<-3$V. If we believe, that this effect comes from parasitic $T$-dependence of the capacitance (Eq.~\ref{cvst}), we can subtract some linear-in-gate voltage dependence from the data, {and obtain the dependence shown by crosses in Fig.~\ref{dsdnlimit}).} 
The obtained $\partial S/\partial p(V_g)$ {dependence} can not be fitted with the theoretical model, as it demonstrates strong positive signal at charge neutrality point, contrary to expected zero value.

We speculate, that this seemingly unphysical value  originates from strong temperature dependence of the spectrum\cite{prl2018,krishtopenko}. Indeed, at the charge neutrality point, E1 band starts to fill. Energy of the bottom of this band is very sensitive to temperature($100 \mu$V/K)\cite{krishtopenko}. As a result, chemical potential for holes starts decreasing $\partial \mu/\partial T<0$, which can explain huge and positive $\partial S/\partial p$ signal at charge neutrality point.

Therefore, we restricted the measurements of the entropy-per-electron to the lowest temperatures.

\section{Appendix II}
{Multi-liquid model is broadly used to extract densities and mobilities in two-dimensional systems\cite{minkov2016, Fletcher2005} , topological insulators\cite{taskin2011,he2012,kuntsevich2016,xiong2013}, semimetals \cite{minkovsemimetal,kvonsemimetal, singh2017, huang2015} etc. This model suggests that in magnetic field carrier densities and mobilities do not depend on the magnitude of the field, and that the conductivity tensor is a sum of those for different carriers:
$$\sigma=\sum_{i=1}^N \frac{p_ie\mu_i}{1+(\mu_i B)^2}\left( \begin{array}{cc}
 1  & \mu_i B \\
-\mu_i B  & 1 \\
\end{array} \right) $$
Here $p_i$ denote carrier densities (in our case we use $p_i$ as the relevant carriers are $p$-type,the same formula is valid for n-type and mixed type carriers), and $\mu_i$ denote carrier mobilities. Resistivity tensor is inverse to the conductivity one $\rho=\sigma^{-1}$.  $p_i$ and $\mu_i$ serve as fitting parameters to describe $\rho_{xx}(B)$ and $\rho_{xy}(B)$ dependencies. 
This model is applicable if (i) mobilities of different groups are not identical (in this case nonlinear $\rho_{xx}(B)$ and Hall coefficient $R_H(B)\equiv\rho_{xy}(B)/B$ dependences emerge in magnetic fields about $1/\mu_{max}$), and (ii) in some range of magnetic fields contributions to the conductivity tensor from different components are comparable.

\begin{figure}
\centerline{\psfig{figure=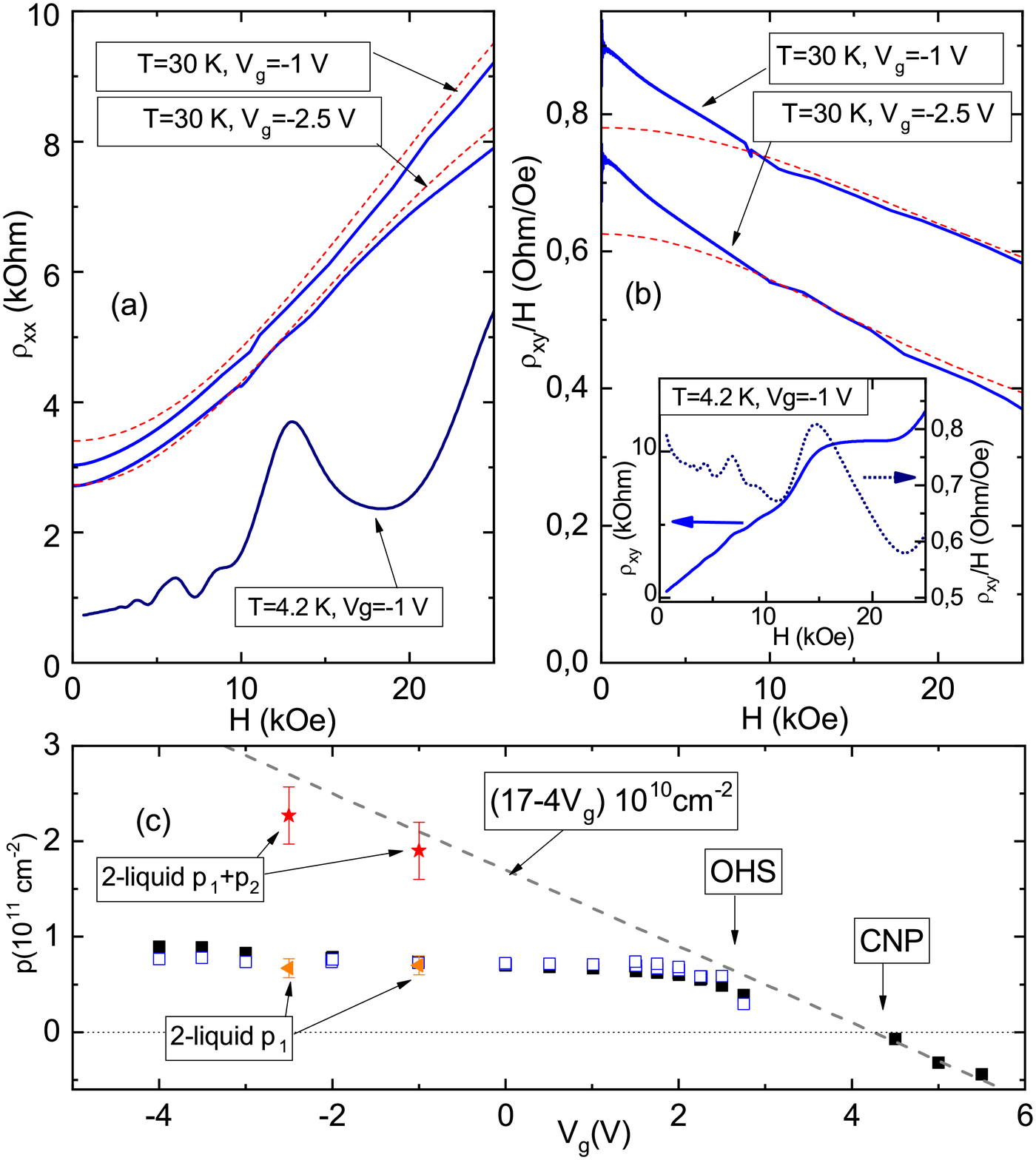,width=250pt}}
\caption{(a) Solid lines- magnetic field dependencies of the resistivity for 110623 sample at 30~K (top two lines) and 4.2K (bottom line). Dashed lines show the two-component fit of the data. (b) Hall coefficient as function of magnetic field for the same sample and gate voltages as in panel (a)(solid lines) and the corresponding theoretical fits (dashed lines). The insert shows Hall resistance (solid line, left axis) and Hall coefficient (dotted line, right axis) at low temperature 4.2~K. The values of the temperatures and gate voltages are indicated in the panels. (c) Comparison of the elevated temperature 2-liquid  fitting parameters from Table~\ref{twoliquid} $p_1$(triangles) and $(p_1+p_2)$(stars), with carrier densities determined from low temperature Hall coefficient (black boxes)and Shubnikov-de Haas oscillations $p_{SdH}$(empty boxes). Low temperature data and extrapolation of total carrier density (dashed line) are taken from Fig.~\ref{fig3}a}.
\label{twoliquid}

\end{figure}

In our case, application of multi-liquid model at low $T$ is hindered by: (i) the presence of three groups of carriers (low-mobility heavy carriers and two groups of light carriers), which requires 6 (!) adjustable fitting parameters; (ii) too high mobility($\sim 10^5$ cm$^2/$Vs) and low effective masses ($\sim 10^{-2} m_0$) of light carriers,which lead to Landau quantization for $H\sim 2$~kOe.
Fig.\ref{twoliquid} shows an example of magnetoresistance (panel (a), the lowest curve) and Hall coefficient (insert to panel (b)) at 4.2K with clear signatures of the quantum Hall effect for $H>10$~kOe. 
Multi-liquid fit is not applicable for this data.

At elevated temperature, mobility of light carriers drops, whereas for heavy carriers it increases. Magneto-oscillations of light carriers become suppressed due to both mobility and temperature factor. Correspondingly, multi-liquid model could be applied. At 30~K, 3-liquid model does not allow to disentangle two species of light carriers (responsible for low-field regime). However, high field domain ($H>10$~kOe) becomes sensitive to low-mobility group of carriers. Therefore, we used 2-liquid model to evaluate the density and mobility of low mobility group of carriers in high field domain(dashed lines in Fig.~\ref{twoliquid}a,b).

We summarize the adjustable parameters ($p_1$, $p_2$, $\mu_1$, and $\mu_2$) of these fits (shown by dashed lines in Fig.\ref{twoliquid}) in the Table~\ref{fitparameters}.

\begin{table}

\caption{Summary of two-liquid model fitting parameters(densities $p_1$,$p_2$ and mobilities $\mu_1$, $\mu_2$) from Fig.~\ref{twoliquid} (sample 110623, $T=30$~K), and also reference value of total carrier density $p_{Tot}$ and SdH density $p_{SdH}$, found from linear extrapolations (lines in Fig.\ref{fig3}b).\label{fitparameters}}
\begin{tabular}{|c|c|c|c|c|c|c|}
  \hline
    V$_g,$& $p_1$, $10^{10}$ & $\mu_1$,&$p_2$,$10^{10}$& $\mu_2$,&$p_{Tot}$,$10^{10}$&$p_{SdH}$,$10^{10}$\\

  V & cm$^{-2}$&cm$^2$/Vs&cm$^{-2}$&cm$^2$/Vs&cm$^{-2}$&cm$^{-2}$\\
  \hline
-1& 7&$5\cdot 10^4$&12&1900&21&7.45\\  
-2.5& 6.7&$2.5\cdot 10^4$&16&1000&27&7.67\\
\hline
\end{tabular}
\end{table}

The sum $p_1+p_2$ (stars in Fig.~\ref{twoliquid}c) is roughly about the total carrier density provided by the gate voltage (dashed line in Fig.~\ref{twoliquid}c). Moreover, $p_1$ value (triangles in Fig.\ref{twoliquid}c) is about the total SdH density
. This agreement suggests (similarly to~ Ref.\cite{minkov2016}) that Fermi-level pinning occurs due to delocalized band states, rather than due to impurity states at the interfaces or in the barriers. Mobility of the heavy carriers is rather small($\sim 1000$ cm$^2/$Vs).  This value is significantly lower than those for $d>8$ nm HgTe QWs (for $p>2\cdot 10^{11}$ cm$^{-2}$ the mobility was $\sim 10^4$~cm$^2$/Vs \cite{minkovVB2017}).

}

{\section{Appendix III}}
The electrical capacitance of the structure is roughly constant across the OHS point. The values $d p_{\rm Total}/dV_g$ in regions A (where only carriers of type 1 are present) and B(where carriers of both types are present) are therefore equal, see Fig.\ref{fig3}a. This condition 
reads:
\begin{equation}
\frac{{dp_{\rm 1}}^A}{dV_g}=\frac{{dp_{\rm 1}}^B}{dV_g}+\frac{{dp_{\rm 2}}^B}{dV_g}
\end{equation}

Since the chemical potential for groups of light (1) and heavy(2) carriers has a common value, the variations of their densities are proportional to the corresponding densities of states $dp_i=D_idE$. From division of the above expression by ${dp_1}^B$ we get:
\begin{equation}
\frac{D_2}{D_{\rm 1}^B}=\frac{dp_{1}^A}{dp_{1}^B}-1
\label{sdhDOS}
\end{equation}

Here we assume that $D_1$ change weakly across the OHS point.

{\section{Appendix IV}}
{ Shubnikov-de Haas oscillations are commonly used to determine subband parameters in two-dimensional systems
\cite{Pudalov2002,Fletcher2005,Chiu2011,minkovVB,Karalic2019,Zhang2001}. In the paper, we did not use a popular method of Fourier transform analysis of SdH\cite{Chiu2011,Karalic2019,Zhang2001}  due to the following reasons. First, the amount of SdH oscillations was not large ($\sim 10$, in higher magnetic field the onset of the quantum Hall effect regime restricts the available field domain). Second, the result of Fourier transform depends strongly on the relations between values of the function at the ends of the interval. Instead, we used direct fitting procedure with a sum of two Lifshits-Kosevich formulas, similarly
to Refs.\cite{minkovVB, Pudalov2002, Fletcher2005}, as explained in the body of the paper.

SdH pattern (Fig.~\ref{massfit}a) noticeably deviates  from the one-component simple Lifshits-Kosevich formula. A fit with two components  agrees well with the data (lines in Fig.~\ref{massfit}a). Amplitudes of each component separately demonstrate systematic dependencies on temperature (shown by vertical arrows in Figs.\ref{massfit}b,c). Moreover, the effective masses, extracted from these dependencies are almost magnetic field-independent (Fig.\ref{massfit}d). This suggests the validity of the fitting procedure. We evaluate an error in effective mass determination as the standard deviation of the mass values in different magnetic fields.

Another important observation from Figs.~\ref{massfit}a-c is that the SdH carrier density for both branches ($p_1$, $p_2$, and, correspondingly, $p_{~Tot}\equiv p_1+p_2$) weakly yet systematically decreases with $T$. This behaviour could be explained. Indeed, close to the OHS point, we observe huge signal $\partial S/\partial p=-\partial \mu/\partial T$. The small decrease of the chemical potential for holes with temperature 
decreases the their density by $\sim \partial \mu/\partial T \times \Delta T\times D_1\approx 0.5 k_B \times 3 K \times 0.05m_0/(2\pi\hbar^2)\sim 10^9$ cm$^{-2}$. This value is in reasonable agreement with the experimentally observed density variation. 
Decrease in density is caused by the increase in the broadening of the Fermi step with $T$. Thus, for higher temperature more states from local valleys become populated, and light subbands deplete.

 This observation again confirms the validity of both the overall picture and SdH fitting procedure.

}
\begin{figure}
\centerline{\psfig{figure=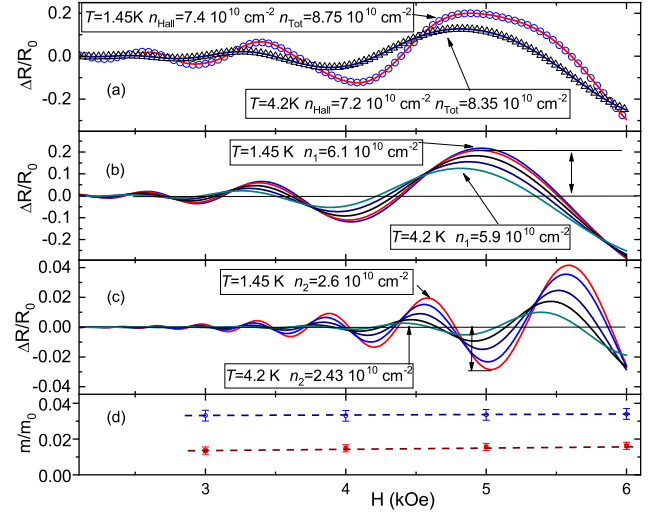,width=250pt}}
\caption{ (a) Example magnetic field dependence of the magneto-oscillations for sample 110623 close to OHS. Symbols are experimental data, lines are fits. (b) and (c) two harmonics at different temperatures from 1.4~K to 4.2~K. (d)Effective masses determined at different magnetic fields.}
\label{massfit}

\end{figure}

\newpage


\begin{thebibliography}{100}
\bibitem{aboutchempot}{Elevated density of states at OHS point(pinning of the chemical potential) might be either due to the states in local valleys (our scenario) or due to some localized states at the heterojunctions or in the barriers. The latter should have extremely low mobility. In Appendix II at elevated temperatures we clearly show that the mobility of these states is high enough, that supports our scenario.}
\end{thebibliography}
\end{document}